\documentclass[prl,floats,twocolumn,showpacs,floatfix,superscriptaddress]{revtex4}
\usepackage{graphics,graphicx,dcolumn,bm,fleqn,epic,eepic,float}
\usepackage{amssymb,amsmath,multirow,rotate,color,float}
\usepackage[latin1]{inputenc}

\newcommand{\ub}{\mbox{{\bf u}}}

\newcommand{\Ub}{\mbox{{\bf U}}}

\newcommand{\Gb}{\mbox{{\bf G}}}

\def\nablab{\mbox{\boldmath$\nabla$\unboldmath}}

\def\nablab{\mbox{\boldmath$\nabla$\unboldmath}}

\def\mean#1{\langle #1 \rangle}

\def\nablab{\mbox{\boldmath$\nabla$\unboldmath}}

\newcommand{\kb}{\mathbf{k}}

\def\mean#1{\langle #1 \rangle}

\begin{document}

\title{ Transverse flow in thin superhydrophobic channels}

\author{Fran\c{c}ois Feuillebois}
\affiliation{LIMSI, UPR 3251 CNRS,
91403 Orsay, France}

\author{Martin Z. Bazant}
 \affiliation{Departments of Mathematics and Chemical Engineering,
Massachusetts Institute
  of Technology, Cambridge, MA 02139 USA}

\author{Olga I. Vinogradova}
 \affiliation{A.N.~Frumkin Institute of Physical
Chemistry and Electrochemistry, Russian Academy of Sciences, 31
Leninsky Prospect, 119991 Moscow, Russia}

\affiliation{ITMC and DWI, RWTH Aachen, Pauwelsstr. 8,
52056 Aachen, Germany}


\date{\today}
%\received{\today}

\begin{abstract}
We provide some general
theoretical results to guide the optimization of transverse
hydrodynamic phenomena in superhydrophobic channels.  Our focus is on the canonical micro- and nanofluidic
geometry of a parallel-plate channel with an arbitrary two-component (low-slip and high-slip) coarse texture, varying on scales larger than the channel thickness. By analyzing rigorous bounds on the permeability, over all possible
patterns, we optimize the area fractions, slip lengths, geometry and orientation of the surface texture to maximize
transverse flow. In the case of two aligned  striped surfaces, very strong transverse flows are possible. Optimized superhydrophobic surfaces may find applications in passive microfluidic mixing and amplification of transverse electrokinetic phenomena.

\end{abstract}
 
\maketitle

{\bf Introduction.--} Hydrophobic solid surfaces with special textures can exhibit greatly enhanced (``super'') properties, compared to analogous flat or slightly disordered surfaces~\cite{quere.d:2005}. If the liquid follows the topological variations of the surface (the Wenzel state), roughness can not only significantly increase hydrophobicity, but can also lead to a giant drop's adhesion. In contrast, when the recessed regions of the texture are filled with gas (the Cassie state), roughness can dramatically lower the ability of drops to stick and produce remarkable liquid mobility. At the macroscopic scale, such surfaces are ``self-cleaning'', causing droplets to roll (rather
than slide) under gravity and rebound (rather than spread) upon
impact. At the microscopic scale, superhydrophobic surfaces could revolutionize microfluidic lab-on-a-chip systems~\cite{stone2004,squires2005}, which are becoming widely used in biotechnology. The large effective slip of superhydrophobic surfaces~\cite{ybert2007,lubrication} compared to simple, smooth channels~\cite{vinogradova.oi:2009,vinogradova.oi:2003,charlaix.e:2005} can greatly lower the viscous drag of thin microchannels and reduce the tendency for clogging or adhesion of suspended analytes. Superhydrophobic surfaces can also amplify electrokinetic pumping or energy conversion in microfluidic devices, if diffuse charge in the liquid extends over the gas regions~
\cite{squires.tm:2008,bahga:2009,Huang08}.

Superhydrophobic surfaces in nature (e.g. the lotus leaf) are typically isotropic, but microfabrication has opened the possibility of highly anisotropic textures~\cite{quere.d:2008}. The effective hydrodynamic slip~\cite{tensor,stone2004,kamrin2010} (or electro-osmotic mobility~\cite{bahga:2009}) of anisotropic textured surfaces is generally {\it tensorial}, due to secondary flows {\it transverse} to the direction of the applied pressure gradient (or electric field~\cite{ajdari2001}).  In the case of grooved no-slip surfaces (Wenzel state), transverse viscous flows have been analyzed for small height variations~\cite{stroock2002a} and thick channels~\cite{wang2003}, and herringbone patterns have been designed to achieve passive chaotic mixing during pressure-driven flow through a microchannel~\cite{stroock2002b,stroock2004,villermaux2008}. Convection is often required to mix large molecules, reagents, or cells in lab-on-a-chip devices, and passive mixing by textured surfaces 
 can be simpler and more robust than mechanical or electrical actuation. In principle, these effects may be amplified by hydrodynamic slip (Cassie state) and large amplitude roughness (Wenzel state), but we are not aware of any prior work.

\begin{figure}
 \includegraphics*[width=0.4\textwidth]{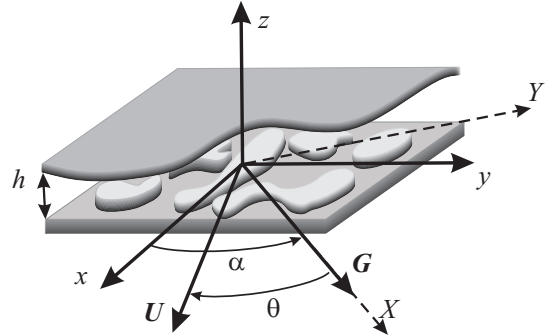}
\caption{Sketch of a thin channel of thickness $h(x,y)$ with notation for directions along the plates. In the Wenzel state the liquid would conform the solid surface, but in the Cassie state it remains free-standing at the top of the roughness}
\label{narrow_channel}
\end{figure}

In this Letter, we present some general theoretical results to guide the optimization of transverse
hydrodynamic phenomena in a thin superhydrophobic channel.  We consider an arbitrary
coarse texture, varying on scales larger than the channel thickness, and optimize its orientation and geometry to maximize pressure-driven transverse flow. Our consideration is based on the theory of heterogeneous porous
materials~\cite{Torquato:2002}, which allows us to derive bounds on transverse flow over all possible
patterns~\cite{lubrication}.

{\bf General considerations.--}
\label{sec:angle}
We consider the pressure-driven flow of a
viscous fluid between two textured parallel plates (``+'' and
``-'') separated by $h (x,y)$, as sketched in Fig.\ref{narrow_channel}. Channel thickness,
$h$, is assumed to vary slowly in directions $x$ and $y$ along the plates. We assume a very general situation, where sectors of different $h$ are characterized by spatially varying, piecewise constant, slip
lengths $b^+(x,y)$ and $b^-(x,y)$.

To evaluate the transverse flow, we calculate the velocity
profile and integrate it across the channel to obtain the
depth-averaged velocity $\ub$ in terms of the pressure gradient
$\nablab p$ along the plates. As usual for the Hele-Shaw cell, the
result may be written as a Darcy law
\begin{equation}
 \ub = - \frac{k}{\mu} \nablab p ,
\label{Darcy}
\end{equation}
where the local permeability is:
\begin{equation}
 k = \frac{h^2}{12} \left[ 1 + \frac{3 (\beta^+
+\beta^-
+4 \beta^+ \beta^-)}{1+\beta^+
+\beta^-} \right]
\label{local_k}
\end{equation}
with
$\beta^+=b^+/h$ and $\beta^-=b^-/h$.
Averaging (\ref{Darcy}) over the heterogeneities in $h,\beta^+,\beta^-$ (see \cite{lubrication} for details), we obtain:
\begin{equation}
 \Ub = - \frac{\kb^*}{\mu} \cdot \nablab P ,
\label{Darcy_averaged}
\end{equation}
where $\Ub, P$ denote the averages of $\ub,p$, respectively.
To simplify the notation, let $\Gb=-(1/\mu)\nablab P$ and $G=|\Gb|$.

A general inhomogeneous medium is characterized by an effective permeability tensor
$\kb^*$ with eigenvalues $k_\|$
along $x$ and $k_\perp$ along $y$, where $k_\| \ge k_\perp
> 0.$ The vector $\Gb$ is applied at an angle $\alpha$ to $x$. Due to inhomogeneity, the velocity
$\Ub$ is generally at an angle $\theta$ ($0\leq\theta\leq\pi/2$) with respect to $\Gb$. Since
$k_\| \ge k_\perp$, it is expected that $\Ub$ will be preferentially in the direction of $x$.
Letting $X$ be the direction of $\Gb$, and $Y$ the perpendicular direction along the plates, we obtain:
\begin{eqnarray*}
 &U_x = k_\|  G  \cos \alpha
\qquad
U_y = k_\perp G \sin \alpha \\
U_X &= U_x \cos \alpha + U_y \sin \alpha = G \, (k_\| \cos^2 \alpha
+ k_\perp \sin^2 \alpha) \\
U_Y &=-U_x \sin \alpha + U_y \sin \alpha = G \, (-k_\| + k_\perp
) \sin \alpha \cos \alpha
\end{eqnarray*}

Our aim is to optimize the texture and the angle $\alpha$, so
that the angle $\theta$ between $\Ub$ and $\Gb$ is maximum
providing the best transverse flow.
That is, $|U_Y/U_X|$ should be maximum (note that $U_Y < 0$ here).
Since at $\alpha=0$ and $\pi/2$, $U_Y=0$, these cases are readily eliminated (The peculiar case of small $|\pi/2-\alpha|$ will be treated separately below).
It is easy to show that
\begin{equation}\label{ratio}
    \left| \frac{U_Y}{U_X} \right| = \frac{(k_\| - k_\perp) \sin \alpha \cos \alpha}{k_\| \cos^2 \alpha +
    k_\perp \sin^2 \alpha} = \frac{(\lambda^2 - 1) t}{\lambda^2 +
    t^2},
\end{equation}
with $\lambda^2 = k_\|/k_\perp \ge 1$ and $t = \tan \alpha$,
is at a maximum if $t=\lambda$. The value of the maximum is
\[
\left| \frac{U_Y}{U_X} \right| = \frac{1}{2} \left(\lambda -
\frac{1}{\lambda} \right)
\]
Therefore, we have transformed our task to optimization of $\lambda$.
\begin{figure}
\includegraphics[width=7cm,clip]{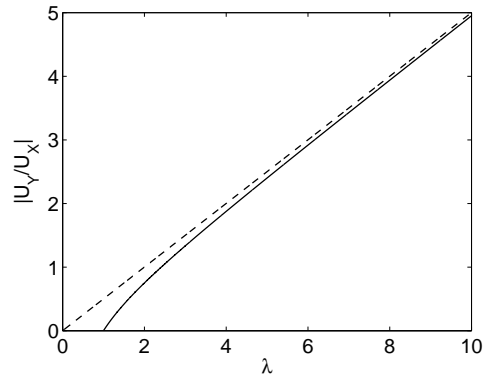}
\caption{The ratio of velocity components, $|U_Y/U_X|$, as a function of $\lambda = \sqrt{ k_\|/k_\perp}$.} \label{fig2}
\end{figure}
To maximize $U_Y/U_X$, $\lambda$
should be as large as possible (Fig.~\ref{fig2}), i.e. $k_\|$ should
be as large, and $k_\perp$ should be as small as possible.

{\bf Two-component medium.--}
\label{sec:thin}
In order to proceed further, the analysis is now restricted to a two-component anisotropic medium with permeabilities $k_1, k_2$. Consider without loss of generality that $0\le k_1 < k_2 < \infty$.
The largest possible $k_\|$ corresponds to the upper Wiener bound~\cite{Torquato:2002}:
\[
 k_\| = \phi_1 k_1 + \phi_2 k_2,
\]
where $\phi_1$ and $\phi_2$ are the area fractions of the
two phases with $\phi_1+\phi_2=1$.
The smallest possible $k_\perp$ corresponds to the lower Wiener bound:
\[
 k_\perp = \left( \frac{\phi_1}{k_1}+\frac{\phi_2}{k_2} \right)^{-1}
\]
A texture satisfying simultaneously both conditions exists: it is a
configuration of stripes.

We then have for this texture
\[
\lambda^2 = \frac{k_\|}{k_\perp} = 1 + \phi_1 (1-\phi_1)
\frac{(k_1 - k_2)^2}{k_1 k_2}
\]
The surface fraction $\phi_1$ corresponding to a maximum of
$U_Y/U_X$ can be found from the equation
\[
\frac{\partial}{\partial \phi_1} \left| \frac{U_Y}{U_X} \right| =
\frac{1}{2} \left(1 + \frac{1}{\lambda^2}\right) \frac{1}{2
\lambda} \frac{\partial \lambda^2}{\partial\phi_1}
\]
which leads to $\partial \lambda^2 / \partial \phi_1 = 0$, which is satisfied at
$\phi_1 = 1/2$. This extremum corresponds to a maximum and then
\[
\lambda^2 = 1 + \frac{(k_1 - k_2)^2}{4 k_1 k_2} =
\frac{(k_1+k_2)^2}{4 k_1 k_2}
\]
Defining:
\[
\overline{k} = \frac{k_1+k_2}{2},\,\,\,
\mean{k} = \sqrt{k_1 k_2},
\]
the maximum occurs at:
\begin{equation}\label{lambda}
    \lambda = \frac{\overline{k}}{\mean{k}}
\end{equation}
and its value is:
\begin{eqnarray} \nonumber
   \left| \frac{U_Y}{U_X} \right| &=& \frac{1}{2} \left( \frac{\overline{k}}{\mean{k}} - \frac{\mean{k}}{\overline{k}} \right) \\
\label{flow}
&=& \frac{1}{2} \left[\frac{1 +
k_2/k_1}{2\sqrt{k_2/k_1}} - \frac{2\sqrt{k_2/k_1}}{1 + k_2/k_1}.
\right]
\end{eqnarray}
which increases monotonously with the anisotropy, $k_2/k_1$. It is interesting to note that the preceding analysis applies to any incompressible, gradient-driven ``flow" in a two-component medium, not only  fluid flow, but also electrical conduction (in which case, we have maximized the transverse current).

{\bf Superhydrophobic channels.--}   We now apply these results to transverse viscous flow in textured, slipping microchannels. We focus first on rough hydrophobic
surfaces in the Cassie state, where trapped air bubbles~\cite{vinogradova.oi:19952,cottin.c:2004,borkent.bm:2007} can lead to dramatic local slip enhancement. To model this, we assume the liquid surface is approximately flat, so the local channel thickness $h=H$ is fixed. This common assumption~\cite{lauga.e:2003,cottin.c:2004} also corresponds to the minimum dissipation~\cite{hyvaluoma.j:2008,ybert2007}. The liquid contacts the solid only over an area
fraction $\phi_1$ of the surface with slip length $b_1$, while the remaining area fraction $\phi_2$ is a free-standing gas-liquid interface. As a simple estimate, lubricating gas sectors of height $\delta$ with
viscosity $\mu_g$ much smaller than that of the liquid
$\mu$~\cite{vinogradova.oi:19951} have a local slip length $b_2 \approx \delta
(\mu/\mu_g)  \approx 50\, \delta$, which can reach tens of $\mu$m. Hydrodynamic slip can also occur at solid  hydrophobic sectors~\cite{vinogradova.oi:1999,lauga2007,bocquet2007}, but with $b_1$ less than tens of nm~\cite{vinogradova.oi:2003,vinogradova.oi:2009,charlaix.e:2005}.

We now consider two cases~\cite{lubrication}:
(I) one slip wall ($\beta^+=\beta; \beta^-=0$), and (II) equal slip on opposite surfaces ($\beta^+=\beta^-=\beta$).
Case (I) is relevant for various setups where the alignment of opposite textures is
inconvenient or difficult. Case (II) is normally used to minimize the drag~\cite{lubrication}. In each case, we have a two-component medium where $\beta$ is either $\beta_1$ or $\beta_2$. The permeability can now be expressed in term of the gap and slip lengths, Eq.~(\ref{local_k}). Then, for $j=1,2$:
\begin{equation}
 k_j =
\begin{cases}
1 + 3\beta_j/[1+\beta_j]       & \mbox{ case (I)}\\
1 + 6 \beta_j  & \mbox{ case (II)}
\end{cases}
\label{def_b_effective}
\end{equation}

Since $h$ is constant, the largest  $k_2/k_1$ obviously corresponds to a
largest physically possible $\beta_2$ with smallest possible $\beta_1$, that is $\beta_1=0$.

\begin{figure}
\includegraphics[width=4cm]{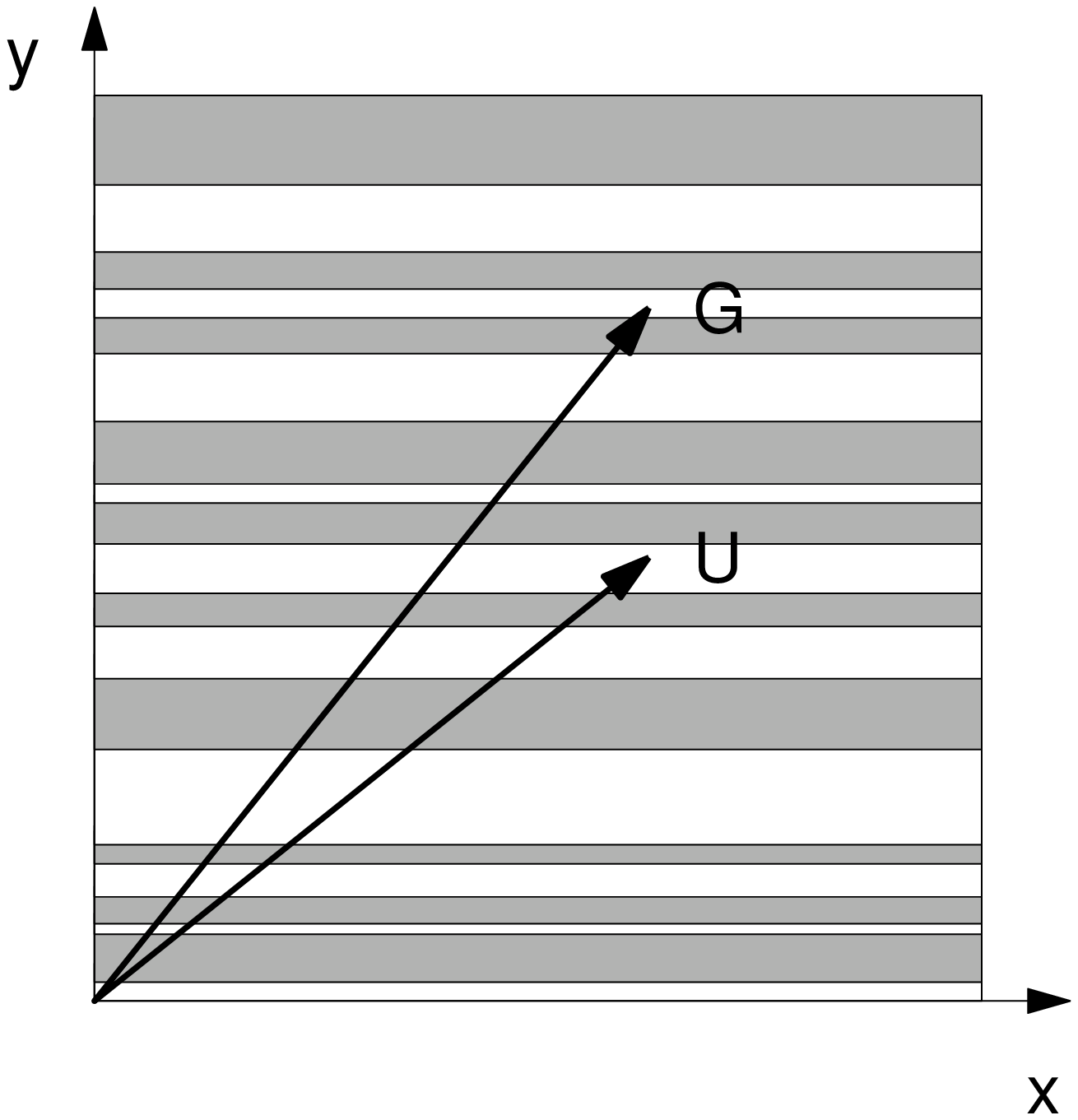}
\includegraphics[width=4cm]{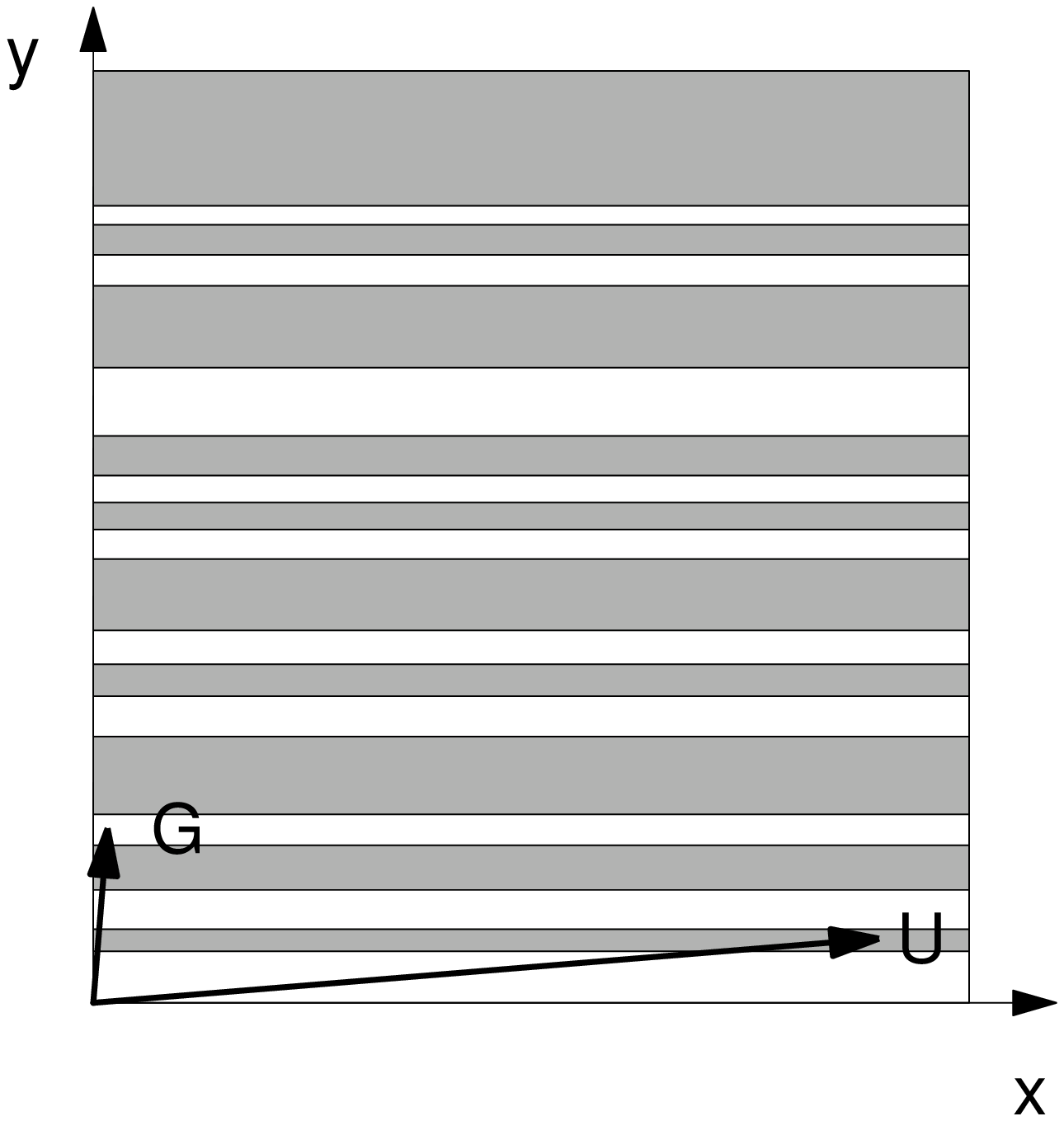}
\caption{Results of transverse flow optimization for flat microchannels with one superhydrophobic and one no-slip wall (case I, left) and two aligned superhydrophobic walls (case II, right).
In both cases, the thickness of stripes is taken as random, to emphasize that the design does not have to be periodic. The mean surface fraction of each component is 1/2. As an example, the distribution of thicknesses is Gaussian with a standard deviation of 0.2. For case II, $\beta_2=100$, and $U_x\simeq 300 G$ has been shortened for visibility.} \label{fig3}
\end{figure}

In case (I), approximating the largest possible $\beta_2$ by $\beta_2\to \infty$ gives $k_2/k_1=4$. We obtain $\lambda = \overline{k}/\mean{k} = 5/4$. The direction of $\Gb$ is $\alpha=\arctan (5/4) = 51.34^{\circ}$. Then $|U_Y/U_X|\to 9/40 = 0.225$. The direction of $\Ub$ is
$\alpha+\theta=\arctan (5/4)-\arctan (9/40) = 38.66^{\circ}$ (see Fig.~\ref{fig3}a), corresponding to a maximum deflection of almost $13^\circ$.

In case (II), the deflection can be more dramatic, but the analysis is more subtle. For $\beta_2\gg 1$, $k_2\simeq 6\beta_2$, the angle $\alpha$ is close to $\pi/2$. Depending on which of the two limits $\beta_2\to\infty$ and $\alpha\to\pi/2$ is taken first, the results are different. The resolution of this singular perturbation problem is to find the {\em significant degeneracy} \cite{Eckhaus:1979}, that is the most general limit from which all other cases may be obtained.
It can be proved here that the significant degeneracy is obtained for our optimum.
We then calculate the following first order approximation:
\begin{eqnarray*}
&& \alpha \simeq \frac{\pi}{2} - \sqrt{\frac{2(1+6\beta_1)}{3\beta_2}}, \\
&&U_X \simeq 4G(1+6\beta_1), \quad
U_Y \simeq -\sqrt{6\beta_2(1+6\beta_1)}\,G.
\end{eqnarray*}
The angle of $\Ub$ and $\Gb$ is then $\theta\simeq \pi/2-2\sqrt{2(1+6\beta_1)/3\beta_2}$.
Note that the flow in this direction close to $\alpha=\pi/2$ is large, but is yet $O(\sqrt{\beta_2})$ smaller than the flow that would exist in the $X$ direction for $\alpha=0$.

For completeness, we also apply our results to the Wenzel state, where the liquid is assumed to follow all the topological variations of the material.  This leads to
a variable thickness for the liquid domain, but fixed hydrodynamic boundary condition. Therefore, the slip length $\beta$ is fixed (possibly zero), but channel thickness $h$ may take two values, $H_1$ and $H_2$ ($H_2>H_1$). It is now convenient to define $H_1=H$ on the top of asperities, as we defined for the Cassie case. It is easy then to show that $k_2/k_1=(H_2/H)^2$. Replacing this value in (\ref{flow}) shows that $H_2/H$ should be as large as possible. This limit is very different from the small surface height modulations and lubrication geometries considered by Stroock et al~\cite{stroock2002b} and suggests that further improvements in passive chaotic mixers may be possible with deeper grooves and thinner channels.

{\bf Concluding remarks.-- } A striking conclusion from our analysis is that the surface textures which optimize transverse flow can significantly differ from those optimizing effective (forward) slip. It is well known that the effective slip of a superhydrophobic surface is maximized by reducing the solid-liquid area fraction $\phi_1$~\cite{ybert2007,lubrication}, until the Cassie state becomes metastable~\cite{quere.d:2008}. In contrast, we have shown that transverse flow in thin channels is maximized by stripes with a rather large solid fraction, $\phi_1=1/2$, where the Cassie state is typically stable. In this situation, the effective slip $\beta^*$ is relatively small~\cite{lubrication}, and yet the flow deflection is very strong (nearly $\pi/2$).

These results may guide the design of superhydrophobic surfaces for robust transverse flows in microfluidic devices. Applications may include flow detection, droplet or particle sorting, or passive mixing. The latter results from interactions with side walls, which produce transverse vortices (due to pressure-driven backflow) and overall helical streamlines, which can be made chaotic for efficient mixing by modulating the surface texture in the axial direction, e.g. with herringbone patterns~\cite{stroock2002b,stroock2002a,stroock2004,villermaux2008}. Compared to the grooved no-slip surfaces with small height variations used in prior work, we have shown that slipping (Cassie) and highly rough (Wenzel) surfaces can exhibit much stronger flow deflection in thin channels, which could lead to more efficient mixing upon spatial modulation of the texture.

Another fruitful direction could be to consider transverse electrokinetic phenomena~\cite{ajdari2001}, e.g. for flow sensors or electro-osmotic pumps~\cite{gitlin2003}. It was recently shown that flat superhydrophobic surfaces can exhibit tensorial electro-osmotic mobility~\cite{bahga:2009}: Anisotropy is maximized if the Debye screening length $\lambda_D$ is comparable to the texture scale, and the gas-liquid interface is uncharged (which, again, does not maximize forward flow); the electro-osmotic mobility scales as, ${\bf M_e} \propto ({\bf I} + {\bf b}/\lambda_D)$ in the limit of thick double layers and even thicker channels ($L \ll \lambda_D \ll H$)~\cite{bahga:2009}. If a similar relation holds for thin channels ($H \ll L \ll \lambda_D$), then our results for the effective ${\bf b}$ (from ${\bf k}$ in case II~\cite{lubrication}) suggest that transverse electrokinetic phenomena could be greatly amplified by using striped superhydrophobic surfaces.

This research was partly supported by the DFG under the Priority programme ``Micro and nanofluidics'' (grant Vi 243/1-3).

%OIV was supported
%by the DFG through its priority programme ``Micro- and
%nanofluidics'' (grant Vi 243/1-3).
\bibliography{slip7}
\end{document}